\documentclass[prl,twocolumn,showpacs,superscriptaddress,amssymb]{revtex4}

\usepackage{psfrag,graphicx}
\usepackage{dcolumn}
\usepackage{bm}
\usepackage{amsfonts,amssymb,amsmath}        % for math symbols.

\usepackage{xcolor}

\begin{document}

\title{Topological effective field theories for Dirac fermions from index theorem} 
 
\author{Giandomenico Palumbo}
\affiliation{School of Physics and Astronomy, University of Leeds, Leeds, LS2 9JT, United Kingdom}

\author{Roberto Catenacci}
\affiliation{Dipartimento di Scienze e Innovazione Tecnologica, Universit\`a del Piemonte Orientale, viale T. Michel 11, 15121 Alessandria, Italy}

\author{Annalisa Marzuoli}
\affiliation{Dipartimento di Matematica ``F. Casorati'', Universit\`a degli Studi di Pavia, via Ferrata 1, 27100 Pavia, Italy and INFN, Sezione di Pavia}

%\date{\today}


\begin{abstract}
Dirac fermions have a central role in high energy physics but it is well known that they 
emerge also as quasiparticles in several condensed matter systems supporting topological order.
We present a general method for deriving the topological effective actions of (3+1) massless Dirac fermions
living on general backgrounds and coupled with vector and axial--vector gauge fields.
The first step of our strategy is standard (in the Hermitian case) and consists in connecting
the determinants of Dirac operators with the corresponding analytical indices 
through the zeta--function regularization. Then, we introduce a suitable splitting of the heat kernel that naturally
selects the purely topological part of the determinant ({\em i.e.} the topological effective action). 
This topological effective action is expressed in terms of gauge fields using
the Atiyah--Singer index theorem which computes the analytical index in topological terms. The main new result of this paper is to provide a  consistent extension of this method to the non Hermitian case where a well--defined
determinant does not exist.
Quantum systems supporting relativistic fermions can thus be topologically classified on the basis of their response to the presence of (external or emergent) gauge fields
through the corresponding topological effective field theories. 
\end{abstract}

\maketitle

\section{Introduction}
The issue of connections between topological quantum field
theories and condensed
matter systems has been intensively investigated
over the years in a variety of different contexts \cite{Zhang1,pachos1, Fradk1}. For example, in two space dimensions the occurrence of 
topological order in ground states of Hamiltonians
\cite{wen} and the existence of quasiparticle excitations
associated with fractional statistics \cite{wilc2} are  crucial features of a variety of many--body microscopic
systems. The existence of such kind of phases of matter represents a  challenge also for theoretical physics because they cannot be described by Landau theory of spontaneous symmetry breaking and thus require more sophisticated mathematical tools.

Besides the case of quantum Hall states which are described by Chern--Simons--type
 topological effective field theories \cite{Zhang2, Fradk2}, topological phases of matter 
 have been recently recognized or conjectured to occur in different physical substrata, such as topological insulators, topological
superconductors, graphene and cold atoms in optical lattices (see {\it e.g.} \cite{moore, qi, marz, pachos}).

In several cases, Dirac fermions emerge as
quasiparticles evolving in the bulk. It is
worth recalling that before the discoveries of graphene \cite{geim} and
topological insulators \cite{molen}, it was not generally thought that
relativistic quantum field theory could  be so relevant in
solid--state physics. However, it is by now generally accepted  that the bulk of topological
insulators and superconductors can be described by gapped Dirac Hamiltonians \cite{schn} 
and that the appearance of relativistic massless particles in the bulk is a hallmark of certain gapless topological phases \cite{volo, kane, ryu0}.

The topological properties together with the relativistic
dynamics of fermions provide strong evidence that the low energy
effective theories can be identified with topological quantum field
theories such as Chern--Simons and BF theories \cite{blau}, 
the latter being defined in both (2+1) and (3+1)--dimensions. 
Indeed, actions of  topological effective field theories (TEFT)
describe  universal global properties
of the physical states and the coefficient of the topological terms can
be identified with the topological order parameter of the system.\\
It is therefore of crucial importance to single out the
specific bosonic field theories corresponding to different microscopic fermionic systems. 
TEFTs emerge through the path integral quantization prescription  by
integrating out the fermionic degrees of freedom in the Dirac
actions with fermions coupled to suitably chosen gauge fields.\\

In this article we present a general method, firstly outlined in \cite{palu}, 
for deriving the topological part of the effective actions of (3+1) massless Dirac fermions
living on general (flat or curved) background manifolds and coupled with vector and axial--vector gauge fields. When an axial--vector field is coupled with fermions, the corresponding Dirac operators are non--Hermitian and we propose a consistent way of defining also in this case a corresponding TEFT. 
Although the total effective action, defined as usual as a regularized determinant, might be 
affected by anomalies, our definition of its topological part (see Eqs. (27)--(35) below)
is gauge and axial invariant.

Our method suggests also a sort of topological classification of these fermionic systems depending on their response to the presence of gauge fields, considered as external or emergent fields \cite{naga}. Such an approach follows the same philosophy proposed in \cite{ryu}, where topological insulators and superconductors have been classified on the basis of quantum anomalies \cite{bert}.

The first step of our strategy is standard (in the Hermitian case) and consists in connecting
the determinants of Dirac operators with the corresponding analytical indices 
through the zeta--function regularization. Then, we introduce a suitable splitting of the heat kernel that naturally
selects the purely topological part of the determinant ({\em i.e.} the topological effective action). 
This topological effective action is expressed in terms of gauge fields using
the Atiyah--Singer index theorem which computes the analytical index in topological terms \cite{naka}.
The main new result of this paper is to provide a consistent extension of this method to the non Hermitian case (where a well--defined determinant does not exist).

It is worth to remark that various types or versions of index theorems have already found direct applications in many condensed matter systems, see {\em e.g.} \cite{stone, wein, fukui}.
 However, we are not going to address here a systematic analysis of  specific
microscopic models which might sustain the proposed topological actions. Rather, in the final section we illustrate a couple of applications. 
The first one is about  three--dimensional topological insulators where massless Dirac fermions emerge in the bulk at quantum critical points. In this regime our results might help in analyzing the response of such kind of substrata 
in the presence of both an electromagnetic and an Abelian axial--vector gauge fields.
In the second final remark it is argued that there might be established a 
2D--3D duality for topological insulators based on a common characterization of their effective actions in terms of BF theories.

\section{Topological effective actions and index theorem}

Effective field theories are fully consistent quantum field
theories valid in a limited energy range described by appropriate
degrees of freedom. In a quantum system composed by fermions and bosons with relativistic dynamics and described by a microscopic action $S$,  the corresponding effective action $S_{\text{\,eff}}$, is usually  obtained by integrating out the fermionic fields

\begin{equation} \label{Dirac}
  e^{-S_{\text{\,eff}}}\,=\,\int \mathcal{D}\,\overline{\psi}\; \mathcal{D}\psi\; e^{-S}.
\end{equation}

 With this formal prescription in mind, we start considering massless Dirac fermions in a (3+1)--dimensional flat spacetime coupled to a generic vector gauge field $A_{\mu}=A_{\mu}^{\alpha}\,T_{\alpha}$, where $\mu$=1,2,3,4 and $T_{\alpha}$ are the generators of the Lie algebra $\mathfrak{g}$ of a compact Lie group $G$,
 $\alpha =1,2,\dots,r$ ($r$= rank ($\mathfrak{g}$)). 
 The action is given by 

\begin{equation} \label{Dirac2}
  S=\int d^{4}x\; \overline{\psi}\;\displaystyle{\not}D\psi=
  \int d^{4}x\;\overline{\psi}\;\gamma^{\mu}(\partial_{\mu}+A_{\mu})\psi,
\end{equation}

where $\gamma^{\mu}$ are the 4$\times$4 Euclidean Dirac matrices
in the chiral basis and $\displaystyle{\not}D$ is the Hermitian Dirac
operator which has the following matrix form

\begin{equation} \label{Dirac3}
  \displaystyle{\not}D \,=\,
 \begin{pmatrix}
 0 &  \displaystyle{\not}D_{-}\\
\displaystyle{\not}D_{+} & 0
\end{pmatrix}\,,  
 \end{equation}

where $\displaystyle{\not}D_{-}=\displaystyle{\not}D_{+}^{\dag}$
and $\displaystyle{\not}D_{+}=\displaystyle{\not}D_{-}^{\dag}$ are
the chiral components of the operator (here the slash complies with the Feynman slash--notation
taken with respect to the Pauli matrices).

Formally, integrating out the fermionic fields, we get

\begin{equation}\label{foreff}
  S_{\text{\,eff}}\,=\, -\log \; \det \displaystyle{\not}D\, :=\,-\frac{1}{2} \;
  \log \; \det \displaystyle{\not}D^{2}\,,
\end{equation}

where $\det \displaystyle{\not}D$ is the determinant of the Dirac
operator and the projectors are given by

\begin{equation}\label{pro1}
  \left(\tfrac{1+\gamma_{5}}{2}\right)\displaystyle{\not}D^{2}=
   \displaystyle{\not}D_{+}^{\dag}\displaystyle{\not}D_{+};
   \; \,\left(\tfrac{1-\gamma_{5}}{2}\right)
   \displaystyle{\not}D^{2}=
   \displaystyle{\not}D_{+}\displaystyle{\not}D_{+}^{\dag} \,\,,
\end{equation}

with $\gamma_{5}=\gamma_{1}\gamma_{2}\gamma_{3}\gamma_{4}$.

 All the physical properties 
 of the quantum system (at one loop)
are captured by the fermionic determinant that needs to be
regularized. In particular, the
zeta--function regularization of  such kind of determinants
can be briefly summarized as follows
\cite{hawk, fursa, dewitt}. 
 Consider a
positive--definite Hermitian second order elliptic operator $L$.
The zeta function associated to $L$ is defined as

\begin{equation}\label{zeta}
  \zeta (s; L)\,=\,\sum_{\lambda} \lambda^{-s}\,=\,\frac{1}{\Gamma(s)}
  \int^{\infty}_{0}dt \; t^{s-1} \text{tr} \; e^{-L t},
\end{equation}

where $\lambda$ are the positive eigenvalues of $L$, $\Gamma(s)$ is the
gamma function, and the zeta function is expressed in terms of
the corresponding heat kernel. 
When the operator has also zero
eigenvalues, it is usual to deform $L$ into $L'=L+\epsilon$, where
$\epsilon$ is a positive number, so that this new operator has only
positive eigenvalues \cite{note1}.
%\footnote{Here the case of operators with a finite (or infinite)  number of negative eigenvalues is not %explicitly addressed, refer to \cite{gilk} for details. Note that a regularized determinant might be %consistently defined also in these cases. }. 
Its determinant is defined  in terms of its zeta function as

\begin{equation}\label{detL1}
  \log \; \det \; L' \, =\, -\frac{d}{ds} \, \zeta (0; L')\,,
\end{equation}

which is related to the Ray--Singer torsion \cite{ray}.
Divergences in the above expression are regularized by using the
$s$--regularized determinant, related  to the zeta function
as follows

\begin{equation}\label{detL2}
  (\log \; \det \; L')_{s} \,=\, -\mu^{2s}\;\Gamma(s) \, \zeta (s; L')\,,
\end{equation}

where $\mu$ is a positive constant of dimension of a mass 
and $Re(s)>2$.

In the case under examination 
$L'=\displaystyle{\not}D^{2}+\epsilon$,  and the limit
$\epsilon\rightarrow 0$ is taken after the regularization to capture the
effective action, namely 

\begin{equation}\label{main}
\begin{split}
 S_{\text{\,eff}} & =\,\lim_{\epsilon\rightarrow 0} \;-\frac{1}{2}(\log \; \det \;
  \displaystyle{\not}D^{2}+\epsilon)_{s}  \\
   & =\lim_{\epsilon\rightarrow 0} \;
  \frac{1}{2}\;\mu^{2s}
  \int^{\infty}
  _{0}dt \; t^{s-1} \text{tr} \; e^{-(\displaystyle{\not}D^{2}+\epsilon)t}\,.
  \end{split}
\end{equation}

Upon manipulating the heat--kernel (by taking into account the formal properties of exponentials and projection operators) one gets

\begin{equation}\label{heat}
\begin{split}
\quad &\text{tr} \; e^{-(\displaystyle{\not}D^{2}+\epsilon)t}\\ 
& =\, \text{tr} \;
\left(\tfrac{1+\gamma_{5}}{2}\right)
   e^{-(\displaystyle{\not}D^{2}+\epsilon)t}
   -\text{tr} \; \left(\tfrac{1-\gamma_{5}}{2}\right)
   e^{-(\displaystyle{\not}D^{2}+\epsilon)t} \\ 
   &  +2\; \text{tr} \; \left(\tfrac{1-\gamma_{5}}{2}\right)
   e^{-(\displaystyle{\not}D^{2}+\epsilon)t}
    \,=\,K_{1}+K_{2},
    \end{split}
\end{equation}
where, due to  relations (\ref{pro1}),

\begin{equation}\label{gian}
K_{1}\,=\,\text{tr} \;
e^{-(\displaystyle{\not}D_{+}^{\dag}\displaystyle{\not}D_{+}
   +\epsilon) t}- \text{tr} \; e^{-(\displaystyle{\not}D_{+}
   \displaystyle{\not}D_{+}^{\dag}+\epsilon) t}
\end{equation}

and

\begin{equation}\label{gian2}
K_{2}\,=\,2\; \text{tr} \; e^{-(\displaystyle{\not}D_{+}
   \displaystyle{\not}D_{+}^{\dag}+\epsilon) t}.
\end{equation}

$K_{1}$ is the difference of two traces and turns out to be connected with the zero--modes of the Dirac operator. For this reason it includes necessarily the contributions due to topological terms while $K_{2}$ includes the non--topological ones. 
Taking into account only $K_{1}$ as given in (\ref{gian}), the  topological part of the effective action (\ref{main}) can be defined as follows

\begin{equation}\label{index1}
\begin{split}
  \quad & S_{\text{\,eff}}^{\text{top}} \,(s)\\  
  & :=\lim_{\epsilon\rightarrow 0}\frac{1}{2}\;\mu^{2s}\Gamma(s)[
  \zeta (s; \displaystyle{\not}D_{+}^{\dag}\displaystyle{\not}D_{+}+\epsilon)-
  \zeta (s; \displaystyle{\not}D_{+}\displaystyle{\not}D_{+}^{\dag}+\epsilon)]\,.
  \end{split}
\end{equation}

Recall that a local formula for the analytical index of an elliptic
operator $Q$  is  given by \cite{gilk, atiya}

\begin{equation}\label{index2}
  \text{ind}\; Q \,=\, \epsilon^{s}\,[
  \zeta (s; Q^{\dag}Q+\epsilon)-
  \zeta (s;
  QQ^{\dag}+\epsilon)]\,.
\end{equation}

The topological part of the regularized effective action can thus be  expressed in terms of 
the analytic index as

\begin{equation}\label{analindex}
S_{\text{\,eff}}^{\text{top}}(s)\,= \,\lim_{\epsilon\rightarrow
0} \;\frac{1}{2}\,\;\mu^{2s} \,\Gamma(s)\, \epsilon^{-s}\; \text{ind}
\,\displaystyle{\not}D_{+} \;.
\end{equation}

Choosing the normalization
 $\mu=\sqrt{c\;\epsilon}$, with  $c$ and $s$, $Re(s)>2$,
arbitrary parameters such that $c^{s}\;\Gamma(s)$ $= -2\pi i$, 
the topological part of the effective action can be defined as

\begin{equation}\label{analindex2}
S_{\text{eff}}^{\text{top}} \,:= \, -i\pi \; \text{ind}
\;\displaystyle{\not}D_{+} \,.
\end{equation}

Note that such  normalization is compatible
with the results found
on  employing different regularization schemes in the case of massive
fermions \cite{hosu}. Indeed, taking a finite value of $\epsilon$ and
considering it as a square mass term ($\epsilon=m^{2}$),
(\ref{analindex}) remains unchanged and the TEFTs obtained from this relation
are in complete agreement with respect to the effective actions in
three space dimensions established in \cite{ryu} by resorting to quantum
anomalies. (Actually this is not surprising since it it well known  that there exists a deep connection between index theorem and quantum anomalies \cite{bert}.)
The arbitrariness in the choice of the normalization
coefficient is simply due to the fact that the original 
path integral (\ref{Dirac}) is  unnormalized.\\

Consider now a generalization of the fermionic action (\ref{Dirac2})  where the fermions are coupled to 
both a vector $A_{\mu}$ and an axial--vector $B_{\mu}$ gauge fields
with value in $\mathfrak{g}$ (Lie--algebra indices $\alpha =1,2,\dots,r$ are hidden)
 
\begin{equation}\label{ferAB}
\begin{split}
  S & = \int d^{4}x\; \overline{\psi}\,\displaystyle{\not}D \,\psi \\
  & =
  \int d^{4}x\;\overline{\psi}\,  \gamma^{\mu}\, (\partial_{\mu}+A_{\mu} \,+\,
  \gamma_{5} \,B_{\mu})\,\psi
  \,.
  \end{split}
\end{equation}

The Dirac operator is not Hermitian in Euclidean signature, namely

\begin{equation}\label{Dirac4}
  \displaystyle{\not}D^{\dag}=\displaystyle{\not}\partial+
  \displaystyle{\not}A-\gamma_{5}\displaystyle{\not}B
  \neq\displaystyle{\not}D.
\end{equation}

The corresponding chiral operators are given by

\begin{equation}\label{chiral}
  \displaystyle{\not}D_{+}=\displaystyle{\not}\partial+
  \displaystyle{\not}A+\displaystyle{\not}B \,; \;\;
  \displaystyle{\not}D_{-}=\displaystyle{\not}\partial+
  \displaystyle{\not}A-\displaystyle{\not}B.
\end{equation}

The operators $\displaystyle{\not}D^{2}$ and 
$(\displaystyle{\not}D^{\dag})^{2}$  do not have a well--defined
determinant. However, the operators
$\displaystyle{\not}D^{\dag}\displaystyle{\not}D$ and
$\displaystyle{\not}D\displaystyle{\not}D^{\dag}$ do have a determinant because they are
Hermitian.

%%%%%%%%%%%%%%%%%%%%%%%%%% PARTE AGGIUNTA
With such remarks in mind, associated effective actions might be hard to define. However, as we are here interested only in  topological contributions,  one can use the fact that reversing the orientation of the background manifold is equivalent to pass from
${\not}D$ to ${\not}D^{\dag}$. Denoting by
$ S_{\text{eff}}^{\uparrow} $ and $ S_{\text{eff}}^{\downarrow} $ respectively the
``effective actions'' defined on the background manifold $M$ (with a fixed orientation) 
and on $\overline{M}$ (with the opposite one), the following definition can be given

\begin{equation} \label{Sarrow}
\begin{split}
  S_{\text{eff}}^{\uparrow}\,+\,
  S_{\text{eff}}^{\downarrow} &
  = \, -\log \, \det \displaystyle{\not}D - 
   \log \, \det \displaystyle{\not}D^{\dag}
  \\ 
  & := -\frac{1}{2}
  \log \; \det \displaystyle{\not}D^{\dag}\displaystyle{\not}D
  - \frac{1}{2}\log \; \det \displaystyle{\not}D\displaystyle{\not}D^{\dag}.
  \end{split}
  \end{equation}
Note that, although the first equality is purely formal,   
the expression in the second row is well defined and necessarily contains
(being invariant under a change of orientation) the topological part of the 
effective action to be evaluated.

%%%%%%%%%%%%%%%ATTENZIONE

%In this case, (\ref{eff}) can be rewritten as follows

%\begin{eqnarray}
  %S_{eff}=-\log \; \det \displaystyle{\not}D= 
  %\frac{1}{2} \log \; \det (\displaystyle{\not}D^{\dag})^{2}
  %\\ \nonumber -\frac{1}{2}
  %\log \; \det \displaystyle{\not}D^{\dag}\displaystyle{\not}D
  %- \frac{1}{2}\log \; \det \displaystyle{\not}D\displaystyle{\not}D^{\dag}
  %\end{eqnarray}

The projectors act as follow

\begin{align}\label{pro3}
  \left(\tfrac{1+\gamma_{5}}{2}\right)\displaystyle{\not}D^{\dag}
  \displaystyle{\not}D &=
  \displaystyle{\not}D_{+}^{\dag}\displaystyle{\not}D_{+}\,;
  \\ \nonumber \left(\tfrac{1-\gamma_{5}}{2}\right)\displaystyle{\not}
  D^{\dag}
  \displaystyle{\not}D &=
  \displaystyle{\not}D_{-}^{\dag}\displaystyle{\not}D_{-}
  \,;\\ \nonumber
 \left(\tfrac{1+\gamma_{5}}{2}\right)\displaystyle{\not}D
  \displaystyle{\not}D^{\dag} &=
  \displaystyle{\not}D_{-}\displaystyle{\not}D_{-}^{\dag}\,;
  \\ \nonumber 
  \left(\tfrac{1-\gamma_{5}}{2}\right)\displaystyle{\not}D
  \displaystyle{\not}D^{\dag} &=
  \displaystyle{\not}D_{+}\displaystyle{\not}D_{+}^{\dag}\,.  
\end{align}

Also in this case we regularize using the zeta--function
prescription, namely

\begin{equation}\label{newreg1}
\begin{split}
 &  -\lim_{\epsilon\rightarrow 0}\tfrac{1}{2}\,
  (\log \; \det \displaystyle{\not}D^{\dag}\displaystyle{\not}D +\epsilon)_{s}
 - \lim_{\epsilon\rightarrow 0} \tfrac{1}{2}\,(\log \; \det
  \displaystyle{\not}D\displaystyle{\not}D^{\dag} +\epsilon)_{s}
 \\ 
 & =\lim_{\epsilon\rightarrow 0}\tfrac{1}{2}\,\mu^{2s}\,  
   \int^{\infty} _{0}dt\; t^{s-1}  \text{tr} \,
  \left[e^{-(\displaystyle{\not}D^{\dag}\displaystyle{\not}D+\epsilon)
  t}+ e^{-(\displaystyle{\not}D\displaystyle{\not}D^{\dag}+\epsilon) t}\right]
  \end{split}
\end{equation}

and

\begin{equation}\label{newreg2}
\begin{split}
\quad & \text{tr} \;
  \left[e^{-(\displaystyle{\not}D^{\dag}\displaystyle{\not}D+\epsilon)
  t}+
  e^{-(\displaystyle{\not}D\displaystyle{\not}D^{\dag}+\epsilon) t}\right]
\\ 
& = \,\text{tr} \;\left(\tfrac{1+\gamma_{5}}{2}\right)
   \left[e^{-(\displaystyle{\not}D^{\dag}\displaystyle{\not}D+\epsilon)
  t}+
  e^{-(\displaystyle{\not}D\displaystyle{\not}D^{\dag}+\epsilon)
  t}\right]
  \\ 
  & - \text{tr} \; \left(\tfrac{1-\gamma_{5}}{2}\right)
   \left[e^{-(\displaystyle{\not}D^{\dag}\displaystyle{\not}D+\epsilon)
  t}+
  e^{-(\displaystyle{\not}D\displaystyle{\not}D^{\dag}+\epsilon) t}\right]
  \\
  & +2\; \text{tr} \; \left(\tfrac{1-\gamma_{5}}{2}\right)
   \left[e^{-(\displaystyle{\not}D^{\dag}\displaystyle{\not}D+\epsilon)
  t}+
  e^{-(\displaystyle{\not}D\displaystyle{\not}D^{\dag}+\epsilon) t}\right]\\
  & = \,K_{1}+K_{2}\,,
   \end{split}
\end{equation}
where, thanks to relations (\ref{pro3}), we
have that
\begin{equation}\label{newreg3}
\begin{split}
K_{1} &= \text{tr} \;
e^{-(\displaystyle{\not}D_{+}^{\dag}\displaystyle{\not}D_{+}
   +\epsilon) t}- tr \; e^{-(\displaystyle{\not}D_{+}
   \displaystyle{\not}D_{+}^{\dag}+\epsilon) t} \\ 
 &  - \text{tr} \;
e^{-(\displaystyle{\not}D_{-}^{\dag}\displaystyle{\not}D_{-}
   +\epsilon) t}+ tr \; e^{-(\displaystyle{\not}D_{-}
   \displaystyle{\not}D_{-}^{\dag}+\epsilon) t}\,;
   \end{split}
\end{equation}

\begin{equation}\label{newreg4}
K_{2}=2\, \text{tr} \; e^{-(\displaystyle{\not}D_{+}
   \displaystyle{\not}D_{+}^{\dag}+\epsilon) t}
   +2\; \text{tr} \; e^{-(\displaystyle{\not}D_{-}^{\dag}
   \displaystyle{\not}D_{-}+\epsilon) t}.
\end{equation}

$K_{1}$ includes the topological terms. 
On the basis of (\ref{index1}) and (\ref{index2}), the topological effective action can be defined as 

\begin{equation} \label{Sfin1}
\begin{split}
 & S_{\,\text{eff}}^{\text{top}} (s)\\ 
  & :=\lim_{\epsilon\rightarrow 0}\frac{1}{2}\;\mu^{2s}
  \Gamma(s)[
  \zeta (s; \displaystyle{\not}D_{+}^{\dag}\displaystyle{\not}D_{+}+
  \epsilon)-
\zeta(s;\displaystyle{\not}D_{+}\displaystyle{\not}D_{+}^{\dag}+\epsilon)\\
& -\zeta (s; \displaystyle{\not}D_{-}^{\dag}\displaystyle{\not}D_{-}+
  \epsilon)
  +\zeta (s; \displaystyle{\not}D_{-}\displaystyle{\not}D_{-}^{\dag}+
  \epsilon)]
\end{split}  
\end{equation}

and finally we get

\begin{equation}\label{Sfin2}
\begin{split}
S_{\,\text{eff}}^{\text{top}} & =\lim_{\epsilon\rightarrow
0}\;\tfrac{1}{2}\;\mu^{2s} \,\Gamma(s)\epsilon^{-s}\;( \text{ind}
\;\displaystyle{\not}D_{+}-\text{ind} \;\displaystyle{\not}D_{-}) \\
& = -i\pi \; (\text{ind} \;\displaystyle{\not}D_{+}-\text{ind} \;
\displaystyle{\not}D_{-})\,,
\end{split}  
\end{equation}

where we have used the same normalization adopted in expression (\ref{analindex2}).\\

Thus we have found that for a non--Hermitian Dirac
operator  a  topological effective action can be defined  as
the difference of the analytic indices of
$\displaystyle{\not}D_{+}$ and $\displaystyle{\not}D_{-}\,$.

\section{Classification of effective field theories}
For an elliptic differential operator on a
compact smooth manifold in even dimension, the Atiyah--Singer index theorem
states
that the analytical index is equal to the topological index \cite{naka}. The
latter is a topological invariant depending only on the fibre
bundles (in the present cases  spinor bundles) living on the
manifold. Most significant situations in applications require
the explicit expression of the topological index
of Dirac operators  in terms of fields on either closed manifolds or compact manifolds with 
(suitable) boundary components. 
Recall that the topological index for a compact manifold with a non--empty boundary
is different from the index for a manifold without
boundary. The Atiyah--Patodi--Singer index theorem indeed
generalizes the Atiyah--Singer index theorem by adding the so--called $\eta$
invariants to the other topological indices \cite{gilk,eguc}. However, being $\eta$ 
a geometric invariant characterizing the boundary, 
it is not involved in the following analysis since we are going to focus only 
onto the topological (bulk) sector of the effective actions.\\

Combining the index theorem with the results of the previous section, 
we give below the explicit forms of a few topological effective field theories 
emerging from microscopic quantum systems in (3+1)--dimensional  (flat or
curved)  spacetimes, where  massless fermions are coupled with non--Abelian gauge fields
(the Abelian analogues being derivable in a straightforward  way). 
In the following $M$ is a four--dimensional smooth orientable compact Riemannian manifold (with or without boundaries) or a (generalized) oriented cilynder, namely
$M = \mathcal{M}^3 \times [0,1]$, with  $\mathcal{M}^3$ a three--dimensional Riemannian compact orientable smooth manifold  (in turns with or without boundaries; moreover $M$ is endowed with the natural product 4-metric). Of course suitable boundary conditions should be 
imposed on the fields according to the specific theories and/or applications of interest.
A complete field--theoretic treatment of such issue in the case of (Abelian) TEFT of the BF--type
can be found in \cite{ABMM}.

\subsection{Fermions coupled with a vector gauge field on a flat spacetime}

The microscopic action is

\begin{equation}\label{Smicro1}
  S\,=\,\int d^{4}x\;
  \overline{\psi}\,\gamma^{\mu}(\partial_{\mu}+A_{\mu})\,\psi\,.
\end{equation}

The topological effective action turns out to be

\begin{equation}\label{chernweil}
S_{\,\text{eff}}^{\text{top}}= -i\pi \; \text{ind}
\;\displaystyle{\not}D_{+}=\frac{i}{8\pi}\int_{M} \text{tr}\; F_{A}\wedge F_{A}\,,
\end{equation}

where $\displaystyle{\not}D_{+}=\displaystyle{\not}\partial+
  \displaystyle{\not}A$, the trace is taken over the gauge indices and $F_{A}=dA+A\wedge A$ is the curvature form \cite{note2}.

\subsection{Fermions on a curved spacetime}

\begin{equation}\label{curved}
  S\,=\,\int d^{4}x\; \det(e)\;
  \overline{\psi}\gamma^{\alpha}e_{\alpha}^{\mu}(\partial_{\mu}+
  \omega_{\mu})\psi,
\end{equation}

where $e_{\alpha}^{\mu}$ are the tetrads and $\omega_{\mu}$ the spin connection \cite{naka}.

\begin{equation}\label{Seff2}
S_{\,\text{eff}}^{\text{top}}= -i\pi \; \text{ind} \;\displaystyle{\not}D_{+}\,=\, -\frac{i}{96\pi}
\int_{M} \text{tr}\; R_{\omega} \wedge R_{\omega}\,,
\end{equation}

where $\displaystyle{\not}D_{+}=\displaystyle{\not}\partial+
\displaystyle{\not}\omega$ and  
$R_{\omega}=d\omega+\omega\wedge \omega$ is the Riemann two--form.\\

\subsection{Fermions coupled with a vector gauge field on a curved spacetime}

\begin{equation}\label{Smicro3}
  S\,=\,\int d^{4}x\; \det(e)\;
  \overline{\psi}\,\gamma^{\alpha}\,e_{\alpha}^{\mu}\,(\partial_{\mu}+A_{\mu}+
  \omega_{\mu})\,\psi\,.
\end{equation}

\begin{equation}\label{Seff3}
\begin{split}
\quad & S_{\,\text{eff}}^{\text{top}}= -i\pi \; ind
\;\displaystyle{\not}D_{+}\\ 
&  =\,\frac{i}{8\pi}\int_{M}tr\; F_{A}\wedge F_{A}-
\frac{i\,(\text{dim}\; \rho)}{192 \pi} \int_{M} tr\; R_{\omega}\wedge R_{\omega}\,,
\end{split}
\end{equation}

where $\displaystyle{\not}D_{+}=\displaystyle{\not}\partial+
  \displaystyle{\not}A+\displaystyle{\not}\omega$ and dim $\rho$ is the dimension of the gauge group
representation.

\subsection{Fermions coupled with vector and axial gauge fields on a flat spacetime} 

\begin{equation}\label{axialvector3}
  S\,=\,\int d^{4}x\;
  \overline{\psi}\,\gamma^{\mu}(\partial_{\mu}+A_{\mu}+
  \gamma_{5}B_{\mu})\,\psi. 
\end{equation}

\begin{equation}\label{axialvector2}
\begin{split}
\quad & S_{\,\text{eff}}^{\text{top}}= -i\pi \; (ind \;\displaystyle{\not}D_{+}- ind
\;\displaystyle{\not}D_{-})\\ 
& =\,\frac{i}{8\pi}\int_{M}tr\; F_{A+B}\wedge F_{A+B}-
\frac{i}{8\pi}\int_{M}tr\; F_{A-B}\wedge F_{A-B} \,,
\end{split}
\end{equation}
where $\displaystyle{\not}D_{+}=\displaystyle{\not}\partial+
  \displaystyle{\not}A+\displaystyle{\not}B$,
  $\displaystyle{\not}D_{-}=\displaystyle{\not}\partial+
  \displaystyle{\not}A-\displaystyle{\not}B$.

\subsection{Fermions coupled with vector and axial gauge fields on a curved spacetime} 

\begin{equation}\label{Smicro5}
  S\,=\,\int d^{4}x\; \det(e)\;
  \overline{\psi}\gamma^{\alpha}e_{\alpha}^{\mu}(\partial_{\mu}+A_{\mu}+
  \gamma_{5}B_{\mu}+\omega_{\mu})\psi\,.
\end{equation}

\begin{equation}\label{Seff5}
\begin{split}
\quad & S_{\,\text{eff}}^{\text{top}}= -i\pi \; (\text{ind} \;\displaystyle{\not}D_{+}-\text{ind}
\;\displaystyle{\not}D_{-})\\ 
& =\frac{i}{8\pi}\int_{M}tr\; F_{A+B}\wedge F_{A+B}-
\frac{i}{8\pi}\int_{M}tr\; F_{A-B}\wedge F_{A-B}\,,
\end{split}
\end{equation}
where $\displaystyle{\not}D_{+}=\displaystyle{\not}\partial+
  \displaystyle{\not}A+\displaystyle{\not}B+\displaystyle{\not}\omega$,
  $\displaystyle{\not}D_{-}=\displaystyle{\not}\partial+
  \displaystyle{\not}A-\displaystyle{\not}B+\displaystyle{\not}\omega$.\\
(Note that here the contributions of the $\omega$--field cancels out.)\\

In the last two cases the definition of the field strength $F$ reads

\begin{equation}\label{FAB}
  F_{A \pm B} \,=\, d (A \pm B) + (A \pm B) \wedge (A \pm B)\,.
\end{equation}

\section{Applications}
In this section we start focusing   on  three--dimensional strong topological
insulators such as $Bi_{1-x}Sb_{x}$ (bismuth antimony) which have a
gapped bulk and gapless surface states  protected by
time--reversal symmetry \cite{hsieh}. This means that they are robust against
non--magnetic impurities/disorder which cannot destroy the
conducting states. These surface states consist of an odd number
of massless Dirac fermions  \cite{qi2}.

We can characterize such  topological insulators through  their response
properties  to an electromagnetic field denoted $a_{\mu}$. Indeed in the bulk this
response can be associated with a topological $\theta$--term 
which (in Lorentzian signature) reads

\begin{equation}\label{axion}
S_{\theta}\,=\,\frac{\theta}{32 \pi^{2}}\int
d^{4}x\;\epsilon^{\mu\nu\sigma\tau}F_{\mu\nu}F_{\sigma\tau}=
\frac{\theta}{8 \pi^{2}}\int F_{a} \wedge F_{a} 
\end{equation}

and is commonly referred to as axion electrodynamics \cite{wilc3, moore, qi}.
For a generic value of $\theta$ the axion term breaks  
time--reversal symmetry $T$ as well as parity $P$. These discrete
symmetries are preserved only for $\theta=\pi$ and $\theta=0$ mod
$2\pi$, which characterize topological insulators and standard
insulators respectively. Since the topological bulk action can be
formulated in terms of a Chern--Simons action on the boundary,
$\theta=\pi$ implies a one--half quantum Hall effect on the
surface \cite{Qi3}

\begin{equation}\label{theta}
S_{CS}\,=\,\frac{\sigma_{xy}}{4\pi}\int d^{3}x\;
\epsilon^{\mu\nu\rho}a_{\mu}\partial_{\nu}a_{\rho}\,,
\end{equation}

where $\sigma_{xy}=\pm 1/2$ (in unit of $e^{2}/h$) is the Hall
conductivity. Time reversal symmetry is broken on the
surface also if is preserved in the bulk, and  thus the boundary
surface becomes insulating, {\em i.e.} Dirac fermions acquire mass.
Conversely, we can say also that a topological axion term
describes the electromagnetic response of topological insulators
when $T$ is broken on the boundary surface.

Three--dimensional massless Dirac points are predicted to exist at the phase
transition between a topological and a normal insulator \cite{mura, hsieh}. In the
alloy $Bi_{1-x}Sb_{x}$ which possesses a rhombohedral crystal
structure, the evolution of its band structure has been
experimentally studied. Changing the $Sb$--concentration $x$,
$Bi_{1-x}Sb_{x}$ passes from a topological insulating state to a
normal insulating one. The quantum critical point is at $x\approx
0.04$, where the gap in the bulk becomes zero and massless
three--dimensional Dirac point is realized.

We argue that also at this critical point the material possesses a
topological behavior thanks to the fact that the electromagnetic
response of the bulk gives rise again to the axion term. Indeed,
nearby the quantum critical point, the tight binding
Hamiltonian can be described in terms of its continuum limit. The effective topological action for
massless Dirac fermions coupled with an electromagnetic field, {\it i.e.} 
the Abelian case of (\ref{chernweil}) in the previous section, is
given exactly by the action (\ref{axion}) with $\theta=\pi$ (with
Lorentzian signature the imaginary unit in front of
(\ref{chernweil}) drops out). We argue that 
in this massless regime a new  topological response might be made manifest.\\

%%ESEMPIO 2

It was pointed out in \cite{moore2} the relevant role of BF theory
in 2 and 3D topological insulators. 
Recall that in two--dimensional topological insulators
quantum spin Hall phase can be realized by means of  a superposition of two
quantum Hall systems with the up- and down--spins having opposite
(effective) magnetic field. A double (2+1) Abelian Chern--Simons theory,
equivalent to an Abelian BF theory, describes consistently the
realization of this quantum phase \cite{bern}.\\ 
On the basis of this remark,  we argue that a (3+1) Abelian BF effective term could be associated 
to the bulk at quantum critical point in three--dimensional topological
insulators when an Abelian axial--vector field $b_{\mu}$
together with the electromagnetic field $a_{\mu}$ are introduced. 
It turns then out that the response of the system is described by 
the corresponding Abelian version of the effective action (\ref{axialvector3}) 
(in Lorentzian signature), namely

\begin{equation}\label{BFex1}
\begin{split}
S_{BF} & =\frac{1}{8\pi}\int (F_{a+b}\wedge F_{a+b}- F_{a-b}\wedge
F_{a-b})\\ 
 & =\frac{1}{2\pi}\int F_{b}\wedge
F_{a}
\end{split}
\end{equation}

which represents a BF term because $F_{b}$ takes exactly
the role of the B--field (a similar result for Dirac materials has been found recently \cite{GosRoy}).
Moreover, the above action  induces  surface contributions
since  it can be expressed as a (2+1)  Abelian BF theory on the
boundary and, by resorting to Stokes' theorem, one gets

\begin{equation}\label{BFex2}
\begin{split}
S_{BF} & =\frac{1}{2\pi}\int_{M} F_{b}\wedge
F_{a}=\frac{1}{2\pi}\int_{M} db\wedge da\\
& =\frac{1}{2\pi}\int_{M} d (b\wedge
da)=\frac{1}{2\pi}\int_{\partial M}b\wedge da \,,
\end{split}
\end{equation}

where $PT$ invariance is manifest. This effective action
and the unbroken time--reversal symmetry are compatible with the
presence of quantum spin Hall states on the surface of the system if we identify $b_{\mu}$ with an external (Zeeman) field coupled to the (edge) spin currents \cite{senthil, mudry}. 
Note that such bulk--boundary correspondence can be interpreted as a sort of 2D--3D duality for topological insulators.\\

Further applications in condensed matter involving the effective actions on curved backgrounds of the previous section should include other microscopic quantum systems such as topological superconductors in connection with thermal Hall effect \cite{ryu, qi4, furu}, Dirac semi--metals and 3D optical lattices.

%\section*{Acknowledgments }


\begin{thebibliography}{25}
\bibitem{note1}  
Here the case of operators with a finite (or infinite)  number of negative eigenvalues is not explicitly addressed, refer to \cite{gilk} for details. Note that a regularized determinant might be consistently defined also in these cases.
\end{thebibliography}
\end{document}